\documentclass[12pt]{article}

\usepackage{jheppub,amsbsy,amssymb,latexsym,amsfonts,amsmath,epsfig,multicol,bbm}

\allowdisplaybreaks

\title{ Exotic symmetry and monodromy equivalence in Schr\"odinger sigma models}

\author{Io Kawaguchi$^{\ast}$\footnote{E-mail:~io@gauge.scphys.kyoto-u.ac.jp} 
and Kentaroh Yoshida$^{\ast}$\footnote{E-mail:~kyoshida@gauge.scphys.kyoto-u.ac.jp}}
\affiliation{$^{\ast}${\it Department of Physics, Kyoto University Kyoto 606-8502, Japan}}

\arxivnumber{1209.4147[hep-th]}

\abstract{
We consider the classical integrable structure of two-dimensional non-linear sigma models with target space 
three-dimensional Schr\"odinger spacetimes. There are the two descriptions to describe the classical dynamics:  
1) the left description based on $SL(2,\mathbb{R})_{\rm L}$ and 2) the right description based on $U(1)_{\rm R}$\,. 
We have shown the $sl(2,{\mathbb R})_{\rm L}$ Yangian and $q$-deformed Poincar\'e algebras associated with them.  
We proceed to argue an infinite-dimensional extension of 
the $q$-deformed Poincar\'e algebra. The corresponding charges are constructed by using a non-local map 
from the flat conserved currents related to the Yangian. 
The {\it exotic} tower structure of the charges is revealed  
by directly computing the classical Poisson brackets. 
Then the monodromy matrices in both descriptions are shown to be gauge-equivalent 
via the relation between the spectral parameters. 
We also give a simple Riemann sphere interpretation of this equivalence. 
}

\keywords{Integrable Field Theory, Sigma Models, AdS-CFT Correspondence}

\begin{document}

\maketitle

\section{Introduction}

In the recent studies of the AdS/CFT correspondence \cite{M,GKP,W}, Schr\"odinger spacetimes \cite{Son, BM} 
have been in the spotlight. The spacetimes are basically null-deformations of the usual AdS spaces  
and the relativistic conformal algebra is broken to a subalgebra called the Schr\"odinger algebra \cite{Sch1,Sch2}.  
Those are considered to be holographic dual 
to non-relativistic (NR) CFTs \cite{Henkel,Nishida} realized 
by using ultra cold atoms in real laboratories. This observation is of great significance 
because it might be possible to test string theory from tabletop experiments 
(for reviews see \cite{Lecture1,Lecture2}). 

\medskip 

To elaborate the AdS/NRCFT correspondence, it is interesting to explore the integrability 
of two-dimensional non-linear sigma models with target space Schr\"odinger spacetimes. 
The integrability in AdS/CFT has played an important role 
in studying the matching of the spectra (for a comprehensive review see \cite{review}).
The integrability yields a powerful tool to examine AdS/NRCFT as well. 

\medskip 

Schr\"odinger spacetimes are represented by non-symmetric and non-reductive cosets \cite{SYY},  
hence the classical integrable structure is far from obvious\footnote{It is well known that 
symmetric coset sigma models in two dimensions  
are always classically integrable. For classic works at an early stage of development, 
see \cite{Luscher1,Luscher2,BIZZ,Bernard-Yangian,MacKay}. For a comprehensive book, see \cite{AAR}.}. 
Although the integrability has not been clarified in more than three dimensions so far, 
the sigma models on three-dimensional Schr\"odinger spacetimes
are shown to be classically integrable by explicitly constructing an infinite set of non-local charges 
and showing that the classical $r/s$-matrices satisfy the extended Yang-Baxter equation 
\cite{KY-Sch}\footnote{
For an earlier argument on the construction of conserved charges based on T-duality, see \cite{ORU}. }. 
Thus the spacetimes are regarded as {\it integrable} deformations of AdS$_3$\,. 
An intensive analysis on other integrable deformations has been done in the recent work \cite{BR}. 

\medskip 

In this paper we proceed our previous analysis on the classical integrable structure of 
two-dimensional non-linear sigma models on three-dimensional Schr\"odinger spacetimes \cite{KY-Sch}. 
There are the two descriptions to describe the classical dynamics:  
1) the left description based on $SL(2,\mathbb{R})_{\rm L}$ and 2) the right description based on $U(1)_{\rm R}$\,. 
We have shown the $sl(2,{\mathbb R})_{\rm L}$ Yangian and $q$-deformed Poincar\'e algebras associated with them 
\cite{KY-Sch}. However, while the Yangian is an infinite-dimensional symmetry, 
the $q$-deformed Poincar\'e symmetry is finite-dimensional.  

\medskip 

The main purpose here is to argue an infinite-dimensional extension of the $q$-deformed Poincar\'e algebra. 
The corresponding charges are constructed by using a non-local map from the flat conserved currents 
related to the Yangian. The {\it exotic} tower structure of the charges is revealed  
by directly computing the classical Poisson brackets. 
Then the monodromy matrices in both descriptions are shown to be gauge-equivalent 
via the relation between the spectral parameters. 
We also give a simple Riemann sphere interpretation of this equivalence.

\medskip 

This paper is organized as follows. In section 2 we introduce the classical action of 
two-dimensional non-linear sigma models on three-dimensional Schr\"odinger spacetimes.  
In section 3 the left description based on $SL(2,{\mathbb R})_{\rm L}$ is revisited. 
We carefully reconsider the flat conserved currents, focusing upon the sign of the current improvement. 
In section 4 our previous analysis on the right description based on $U(1)_{\rm R}$ further proceeds. 
An infinite-dimensional extension of $q$-deformed Poincar\'e algebra is shown directly 
by constructing the conserved charges. The tower structure of the charges is generated 
in an {\it exotic} way. Then we show the gauge-equivalence of the monodromy matrices 
in the two descriptions and give a simple Riemann sphere interpretation of it.
Section 5 is devoted to conclusion and discussion. 
In appendix A we show the right current algebra 
that is used to compute the algebra of non-local conserved charges in section 4. 
In appendix B we explain in detail the prescription utilized in computing the algebra of 
an infinite dimensional extension of the $q$-deformed Poincar\'e algebra. 

\section{Setup}

In this section we introduce the classical action of two-dimensional non-linear sigma models 
on three-dimensional Schr\"odinger spacetimes

\subsection{Schr\"odinger spacetimes}

Schr\"odinger spacetimes in three dimensions are known as null-like deformations of AdS$_3$\,. 
And the metric of the spacetimes is given by 
\begin{eqnarray}
ds^2=L^2\left[-2{\rm e}^{-2\rho}dudv+d\rho^2-C{\rm e}^{-4\rho}dv^2\right]\,. 
\label{Sch3-angle}
\end{eqnarray}
The deformation is measured by a real constant parameter $C$\,. When $C=0$\,, the metric (\ref{Sch3-angle}) 
describes the AdS$_3$ space with the curvature radius $L$ and the isometry is 
$SO(2,2)=SL(2,{\mathbb R})_{\rm L}\times SL(2,{\mathbb R})_{\rm R}$\,.  
When $C\neq 0$\,, the $SL(2,{\mathbb R})_{\rm R}$ symmetry is broken to $U(1)_{\rm R}$ while 
the $SL(2,{\mathbb R})_{\rm L}$ symmetry is preserved. Hence the symmetry that survives the deformation is 
$SL(2,{\mathbb R})_{\rm L}\times U(1)_{\rm R}$ in total. This symmetry yields two distinct descriptions to 
describe the classical dynamics as we will see later. 

\medskip 

It is useful to rewrite the metric (\ref{Sch3-angle}) in terms of an $SL(2,{\mathbb R})$ group element $g$,   
\begin{eqnarray}
g={\rm e}^{2vT^+}{\rm e}^{2\rho T^2}{\rm e}^{2uT^-}\,, \qquad 
T^\pm \equiv \frac{1}{\sqrt{2}}\left(T^0\pm T^1\right)\,.
\end{eqnarray}
Here the ${\it sl}(2,{\mathbb R})$ generators $T^a~(a=0,1,2)$ satisfy  
\begin{eqnarray}
\left[T^a,T^b\right]=\varepsilon^{ab}_{~~c}T^c\,,\quad {\rm Tr}\left(T^aT^b\right)=\frac{1}{2}\gamma^{ab}\,, 
\end{eqnarray}
where $\varepsilon^{ab}_{~~c}$ is the anti-symmetric tensor normalized as 
\[
\varepsilon^{01}_{~~\,2}=+1\,, \qquad \gamma^{ab}={\rm diag}(-1,+1,+1)\,.
\]
In the following discussion, 
$\gamma^{ab}$ and its inverse are used to rising and lowering the $sl(2,{\mathbb R})$ indices. 
It is useful to list the light-cone components of $\varepsilon^{ab}_{~~c}$ and $\gamma_{ab}$~: 
\begin{eqnarray}
&&\varepsilon^{-+}_{~~~~2}=+1\,, 
\quad \gamma_{-+}=\gamma_{+-}=-1\,, \quad \gamma_{22}=+1\,. \nonumber
\end{eqnarray}
For definiteness we take a representation of the ${\it sl}(2,{\mathbb R})$ generators,  
\begin{eqnarray}
T^0=\frac{i}{2}\sigma_2\,,\quad T^1=\frac{1}{2}\sigma_1\,,\quad T^2=\frac{1}{2}\sigma_3\,, 
\end{eqnarray}
where $\sigma_i~(i=1,2,3)$ are the Pauli matrices and $T^2$ is taken as the Cartan generator.  

\medskip 

With the group element $g$\,, the left-invariant one-form $J$ is defined as 
\begin{eqnarray}
J \equiv g^{-1}dg\,. 
\end{eqnarray}
Now $J$ is expanded in terms of the ${\it sl}(2,{\mathbb R})$ generators like 
\begin{eqnarray}
J &=& -T^0J^0+T^1J^1+T^2J^2 \nonumber \\ 
&=& -T^+J^--T^-J^++T^2J^2\,, \nonumber 
\end{eqnarray} 
where each of $J^a$'s is written in a simple form, 
\begin{eqnarray}
J^a=2{\rm Tr}(T^aJ)\,.
\end{eqnarray}

\medskip 

It is a turn to rewrite the metric (\ref{Sch3-angle}) in terms of $J$ like 
\begin{eqnarray}
ds^2&=&\frac{L^2}{2}\left[{\rm Tr}(J^2)-2C\left({\rm Tr}[T^-J]\right)^2\right] \nonumber \\
&=&\frac{L^2}{4}\left[-2J^+J^-+(J^2)^2-C(J^-)^2\right]\,. \label{last}
\end{eqnarray}
From the last expression (\ref{last})\,, the Schr\"odinger spacetime can be regarded as a null-deformation of AdS$_3$\,. 

\medskip 

Then the isometry $SL(2,{\mathbb R})_{\rm L}\times U(1)_{\rm R}$ is represented by 
\begin{eqnarray}
g'= {\rm e}^{\beta^a T^a} \cdot g \cdot{\rm e}^{-\alpha T^-}\,, \nonumber 
\end{eqnarray}
where $\beta^a~(a=0,1,2)$ and $\alpha$ are real constant parameters.  
Its infinitesimal form is 
\begin{eqnarray}
\delta^a_Lg=\epsilon_{\rm L}^a\, T^a\, g\,, \qquad \delta^-_Rg = - \epsilon_{\rm R}\, g\,T^-\,. \nonumber  
\end{eqnarray} 
Here the summation is not taken for the repeated indices in the first equation.

\subsection{Schr\"odinger sigma models}

Our purpose is to investigate the classical integrable structure of two-dimensional non-linear sigma models 
on three-dimensional Schr\"odinger spacetime. For simplicity, we shall refer to these sigma models 
as {\it Schr\"odinger sigma models}. 

\medskip 

With the metric (\ref{last}), the classical action is given by 
\begin{eqnarray}
S&=&-\int^\infty_{-\infty}\!\!\!dt\!\int^\infty_{-\infty}\!\!\!dx\,\eta^{\mu\nu}\left[{\rm Tr}\left(J_{\mu}J_{\nu}\right)
-2C{\rm Tr}\left(T^{-}J_{\mu}\right){\rm Tr}\left(T^{-}J_{\nu}\right)\right]\,. 
\label{action}
\end{eqnarray}
The base space is a two-dimensional Minkowski spacetime with the coordinates $x^\mu=(t,x)$ 
and the metric $\eta_{\mu\nu}=(-1,+1)$\,. 
The rapidly dumping boundary condition is taken so that 
the group-valued field $g(x)$ approaches a constant element $g_{\infty}$ at spatial infinities:
\begin{eqnarray}
g(x) ~~\to~~ g_\infty \qquad (x\to\pm\infty)\,. \nonumber 
\end{eqnarray}
That is, $J_{\mu}$ vanishes very rapidly at spatial infinities. 

\medskip 

This setup is not appropriate in considering some applications to string theory.  
However, it is suitable to study infinite-dimensional symmetries generated by an infinite set of 
non-local charges in a well-defined way. The Virasoro constraints are also not taken into account. 

\medskip 

Taking a variation of the action (\ref{action}) leads to the equations of motion,  
\begin{eqnarray}
\partial^{\mu}J_{\mu}-2C{\rm Tr}\left(T^{-}\partial^{\mu}J_{\mu}\right)T^{-}
-2C{\rm Tr}\left(T^{-}J_{\mu}\right)\left[J^{\mu},T^{-}\right]=0\,. 
\label{eom}
\end{eqnarray}
By multiplying $T^-$ and taking the trace, the conservation law of the $U(1)_{\rm R}$ current is obtained as 
\begin{eqnarray}
\partial^{\mu}J^{-}_{\mu} = 0\,. \label{cons-u(1)}
\end{eqnarray}
With this conservation law (\ref{cons-u(1)})\,, the equations of motion in (\ref{eom}) are simplified as 
\begin{eqnarray}
\partial^{\mu}J_{\mu} -2C{\rm Tr}\left(T^{-}J_{\mu}\right)\left[J^{\mu},T^{-}\right]=0\,. 
\label{eom2}
\end{eqnarray}
When $C=0$, the Schr\"odinger sigma models become $SL(2,{\mathbb R})$ principal chiral models. 

\medskip 

The Schr\"odinger sigma models are classically integrable \cite{KY-Sch}. 
The $SL(2,{\mathbb R})_{\rm L}$ and $U(1)_{\rm R}$ symmetries give rise to two descriptions to 
describe the classical dynamics. One is the left description based on $SL(2,{\mathbb R})_{\rm L}$\,.  
The other is the right description based on $U(1)_{\rm R}$\,. For each of them, Lax pairs and monodromy matrices 
are constructed. Then all Lax pairs lead to the identical equations of motion in (\ref{eom2}). 
The classical integrable structure is similar to the hybrid one in the squashed S$^3$ 
and warped AdS$_3$ cases \cite{KYhybrid,KMY-QAA,KMY-monodromy,KY-summary}.

\section{The left description}

One way to describe the classical dynamics is the left description based on $SL(2,{\mathbb R})_{\rm L}$\,. 
A pair of flat conserved currents is obtained by improving the $SL(2,{\mathbb R})_{\rm L}$ Noether current 
appropriately. Then the corresponding Lax pairs and monodromy matrices are constructed in the usual way. 
Both of them lead to the same equations of motion in (\ref{eom2}) and two copies of the Yangians.

\medskip 

This section is mainly a brief review of the previous work \cite{KY-Sch}. 
However, the sign of the improvement is carefully reconsidered here 
(while a specific sign has been considered in \cite{KY-Sch})  
and it will be important in our later discussion.

\subsection{Flat conserved currents and Yangians}

The key ingredient here is the $SL(2,{\mathbb R})_{\rm L}$ conserved current,   
\begin{eqnarray}
j^L_\mu=gJ_\mu g^{-1}-2C{\rm Tr}\left(T^-J_\mu\right)gT^-g^{-1}+\epsilon_{\mu\nu}\partial^\nu f\,. 
\end{eqnarray}
where $\epsilon_{\mu\nu}$ is an anti-symmetric tensor normalized as $\epsilon_{tx}=+1$ 
and $f$ is an undetermined function. The first two terms are obtained by following the Noether's procedure 
and the last term is an improvement term. 

\medskip 

An important point here is that the function $f$ can be taken as 
\begin{eqnarray}
f=-AgT^-g^{-1}\,, \nonumber
\end{eqnarray}
so that the improved current satisfies the following equation, 
\begin{eqnarray}
\epsilon^{\mu\nu}\left(\partial_\mu j^L_\nu -j^L_\mu j^L_\nu\right) 
= \left(A^2-C\right)\epsilon_{\mu\nu}\partial^\mu(gT^-g^{-1})\partial^\nu(gT^-g^{-1})\,. 
\end{eqnarray}
Thus, by choosing the appropriate values 
\begin{eqnarray}
A=\pm\sqrt{C}\,, 
\end{eqnarray}
the improved current satisfies the flatness condition, 
\begin{eqnarray}
\epsilon^{\mu\nu}\left(\partial_\mu j^L_\nu -j^L_\mu j^L_\nu\right)=0\,. 
\end{eqnarray}
The explicit form of the resulting flat conserved currents are given by
\begin{eqnarray}
j^{L_\pm}_\mu=gJ_\mu g^{-1}-2C{\rm Tr}\left(T^-J_\mu\right)gT^-g^{-1} 
\mp \sqrt{C}\epsilon_{\mu\nu}\partial^\nu\left(gT^-g^{-1}\right)\,. 
\end{eqnarray}
The subscript of $L$ ($+$ or $-$) denotes the sign of the improvement term. 
Although only $j_{\mu}^{L_+}$ has been discussed in \cite{KY-Sch}, 
both currents are important in the present case\footnote{The current improvement leads to 
the flat conserved currents also in the squashed S$^3$ case \cite{BFP,KY}. 
This is the case even if the Wess-Zumino term has been added \cite{KOY}.}.   

\medskip 

With the improved currents, an infinite set of conserved non-local charges are recursively constructed 
by following \cite{BIZZ}. The first three charges are given by 
\begin{eqnarray}
Q^L_{(0)} &=&\int^\infty_{-\infty}\!\!\!dx~j^{L_\pm}_t(x)\,, \nonumber \\
Q^L_{(1)} &=&\frac{1}{4}\int^\infty_{-\infty}\!\!\!dx\int^\infty_{-\infty}\!\!\!dy~\epsilon(x-y)\left[j^{L_\pm}_t(x),j^{L_\pm}_t(y)\right]
-\int^\infty_{-\infty}\!\!\!dx~j^{L_\pm}_x(x)\,, \nonumber \\
Q^L_{(2)} &=&\frac{1}{12}\int^\infty_{-\infty}\!\!\!dx\int^\infty_{-\infty}\!\!\!dy\int^\infty_{-\infty}\!\!\!dz~\epsilon(x-y)\epsilon(x-z)\left[\left[j^{L_\pm}_t(x),j^{L_\pm}_t(y)\right],j^{L_\pm}_t(z)\right] \nonumber \\
&&-\frac{1}{2}\int^\infty_{-\infty}\!\!\!dx\int^\infty_{-\infty}\!\!\!dy~\epsilon(x-y)\left[j^{L_\pm}_t(x),j^{L_\pm}_x(y)\right]
+\int^\infty_{-\infty}\!\!\!dx~j^{L_\pm}_t(x)\,, \nonumber \\ 
&& \vdots 
\label{left-bizz}
\end{eqnarray}
where $\epsilon(x-y) \equiv \theta(x-y) - \theta(y-x)$ and $\theta(x)$ is a step function. 
The subscript $(n)$ denotes the degree of non-locality. 
Note that the sign of the improvement term is not relevant at the charge level 
because only $C$ appears in the charges but $\sqrt{C}$ does not. 
Thus the subscript $\pm$ for the charges has been ignored above.

\medskip  

Then the current algebra is given in terms of the classical Poisson bracket $\{~.~\}_{\rm P}$\,, 
\begin{eqnarray}
\left\{j^{L_\pm,a}_t(x),j^{L_\pm,b}_t(y)\right\}_{\rm P} 
&=& \varepsilon^{ab}_{~~c}j^{L_\pm,c}_t(x)\delta(x-y)\,, \nonumber \\
\left\{j^{L_\pm,a}_t(x),j^{L_\pm,b}_x(y)\right\}_{\rm P} 
&=& \varepsilon^{ab}_{~~c}j^{L_\pm,c}_x(x)\delta(x-y)+\gamma^{ab}\partial_x\delta(x-y)\,, \label{ca} \\
\left\{j^{L_\pm,a}_x(x),j^{L_\pm,b}_x(y)\right\}_{\rm P} 
&=& 0\,. \nonumber 
\end{eqnarray}
Here the components of $j^{L_\pm}_{\mu}$ are defined as  
\[
j^{L_\pm,a}_\mu \equiv 2{\rm Tr}\left(T^aj^{L_\pm}_\mu\right)\,. 
\] 
Note that the current algebra (\ref{ca}) does not contain the deformation parameter $C$\,, 
in comparison to the squashed S$^3$ and warped AdS$_3$ cases. 
Thus in the present case, 
redfollowing a prescription of \cite{MacKay}\footnote{
There is a subtlety in computing the Yangian algebra because the current algebra (\ref{ca}) 
contains non-ultra local terms. This is the case even in principal chiral models.
A possible resolution is to take the order of limits 
so as to maintain the Serre relations \cite{MacKay}. 
Actually, we followed this prescription in the previous work \cite{KY-Sch}. 
For recent progress on the issue of non-ultra local terms, see \cite{DMV}.}
, one can show that 
the charges (\ref{left-bizz}) satisfy the defining relations of ${\it sl}(2,{\mathbb R})$ Yangian 
in the sense of the first realization by Drinfeld \cite{Drinfeld}

\subsection{Lax pairs and monodromy matrices}

Let us next consider the Lax pairs. Since the flat conserved currents have already been obtained, 
it is easy to construct the Lax pairs,  
\begin{eqnarray}
L^{L_\pm}_\mu(x;\lambda_{L_\pm}) &=& 
\frac{1}{1-\lambda_{L_\pm}^2}\left(j^{L_\pm}_\mu(x)
-\lambda_{L_\pm}\epsilon_{\mu\nu}j^{L_\pm,\nu}(x)\right)\,.
\label{left lax}
\end{eqnarray}
Here the spectral parameters $\lambda_{L_{\pm}}$ are constant complex numbers.  
Then the commutation relations 
\begin{eqnarray}
\left[\partial_t-L_t^{L_\pm}(x;\lambda_{L_\pm}),\partial_x-L^{L_\pm}_x(x;\lambda_{L_\pm})\right]=0
\end{eqnarray}
give rise to the flatness conditions and the conservation laws for the improved currents. 
We will refer to (\ref{left lax}) as the left Lax pairs.  

\medskip 

The flat conserved currents enable us to construct the monodromy matrices,  
\begin{eqnarray}
U^{L_\pm}(\lambda_{L_\pm})=
{\rm P}\exp\left[\int^\infty_{-\infty}\!\!\!dx~L^{L_\pm}_x(x;\lambda_{L_\pm})\right]\,, 
\label{left-monodromy}
\end{eqnarray}
where the symbol P denotes a path-ordering operation. It is straightforward to show that 
the monodromy matrices are conserved quantities, 
\begin{eqnarray}
\frac{d}{dt}U^{L_\pm}(\lambda_{L_\pm})=0\,.  \nonumber 
\end{eqnarray}
Thus, by expanding (\ref{left-monodromy}) around a certain value of $\lambda_{L_{\pm}}$\,, 
an infinite set of conserved charges are obtained. 
For instance, the $sl(2,{\mathbb R})_{\rm L}$ Yangian generators listed in (\ref{left-bizz}) are reproduced by expanding 
(\ref{left-monodromy}) like 
\begin{eqnarray}
U^{L_\pm}(\lambda_{L_\pm})=\exp\left[\sum_{n=0}^\infty\lambda_{L_{\pm}}^{-n-1}Q^L_{(n)}\right] \nonumber 
\end{eqnarray}
around $\lambda_{L_\pm}=\infty$\,. 

\medskip 

By evaluating the following Poisson brackets \cite{Maillet}\,,
\begin{eqnarray}
&&\left\{L^{L_\pm}_x(x;\lambda_{L_\pm})\stackrel{\otimes}{,}L^{L_\pm}_x(y;\mu_{L_\pm})\right\}_{\rm P} \nonumber \\
&=& \left[r^{L_\pm}(\lambda_{L_\pm},\mu_{L_\pm}),L^{L_\pm}_x(x;\lambda_{L_\pm})\otimes 1+1\otimes L^{L_\pm}_x(y;\mu_{L_\pm})\right]\delta(x-y) \nonumber \\
&& -\left[s^{L_\pm}(\lambda_{L_\pm},\mu_{L_\pm}),L^{L_\pm}_x(x;\lambda_{L_\pm})\otimes 1-1\otimes L^{L_\pm}_x(y;\mu_{L_\pm})\right]\delta(x-y) \nonumber \\
&& -2s^{L_\pm}(\lambda_{L_\pm},\mu_{L_\pm})\partial_x\delta(x-y)\,, \nonumber 
\end{eqnarray}
we obtain the following classical $r$/$s$-matrices 
\begin{eqnarray}
&&r^{L_\pm}(\lambda_{L_\pm},\mu_{L_\pm}) = \frac{h(\lambda_{L_\pm})+h(\mu_{L_\pm})}{2\left(\lambda_{L_\pm}-\mu_{L_\pm}\right)}
\left(- T^+\otimes T^- - T^-\otimes T^+ + T^2\otimes T^2\right)\,, \nonumber \\
&&s^{L_{\pm}}(\lambda_{L_\pm},\mu_{L_\pm}) = \frac{h(\lambda_{L_\pm})-h(\mu_{L_\pm})}{2\left(\lambda_{L_\pm}-\mu_{L_\pm}\right)}
\left(- T^+\otimes T^- - T^-\otimes T^+ + T^2\otimes T^2\right)\,. \nonumber 
\end{eqnarray}
Here we have introduced a scalar function,
\begin{eqnarray}
h(\lambda) \equiv \frac{\lambda^2}{1-\lambda^2} \label{h-function}
\end{eqnarray}
The classical $r/s$-matrices satisfy the extended classical Yang-Baxter equation \cite{KY-Sch}. 

\medskip 
  
Note that the classification of classical $r$-matrix by Belavin-Drinfeld \cite{BD} 
does not make sense for the extended Yang-Baxter equation, but only for the original Yang-Baxter equation. 
Actually, the $r$-matrix cannot be recast into the skew symmetric form with respect to 
the difference of spectral parameters\footnote{We would like to thank B.~Vicedo for this point.}. 

\section{The right description}

The other way to describe the classical dynamics is the right description based on $U(1)_{\rm R}$\,. 
An infinite-dimensional symmetry should be related to this description. But it has not been clarified 
in the previous work \cite{KY-Sch}, 
though the finite-dimensional part is found to be $q$-deformed Poincar\'e algebra. 

\medskip 

Hence, after giving a short review on the previous result, 
we first consider its infinite-dimensional extension. By constructing the conserved charges explicitly, 
the {\it exotic} tower structure of the conserved charges will be revealed. 
Then we construct the associated Lax pairs and monodromy matrices, 
which are gauge-equivalent to the ones in the left description. 
We also give a simple Riemann sphere description of this equivalence.

\subsection{$q$-deformed Poincar\'e algebra}

The conserved current related to the $U(1)_{\rm R}$ symmetry is given by 
\begin{eqnarray}
j^{R,-}_\mu=-J^-_\mu\,.
\end{eqnarray}
For the broken components of $SL(2,\mathbb{R})_{\rm R}$\,, the following non-local currents 
\begin{eqnarray}
&&j^{R,2}_\mu=-{\rm e}^{\sqrt{C}\chi}\left(J^2_\mu+\sqrt{C}\epsilon_{\mu\nu}J^{-,\nu}\right)\,, \label{NL-right1}\\
&&j^{R,+}_\mu=-{\rm e}^{\sqrt{C}\chi}\left(J^+_\mu+\sqrt{C}\epsilon_{\mu\nu}J^{2,\nu}+CJ^-_\mu\right) \nonumber
\end{eqnarray}
are conserved \cite{KY-Sch}. Here we have introduced a non-local field $\chi(x)$ defined as 
\begin{eqnarray}
\chi(x) \equiv -\frac{1}{2}\int^\infty_{-\infty}\!\!\!dy~\epsilon(x-y)\,j^{R,-}_t(y)\,. 
\label{chi}
\end{eqnarray}
Thus the conserved charges are  
\begin{eqnarray}
Q^{R,-}=\int^\infty_{-\infty}\!\!\!dx~j^{R,-}_t(x)\,, \quad
Q^{R,2}=\int^\infty_{-\infty}\!\!\!dx~j^{R,2}_t(x)\,, \quad
Q^{R,+}=\int^\infty_{-\infty}\!\!\!dx~j^{R,+}_t(x)\,. \label{q-Poincare charge}
\end{eqnarray}
Note that the last two charges are non-local due to the presence of (\ref{chi}).   

\medskip 

Then the Poisson brackets of the charges are computed as \cite{KY-Sch} 
\begin{eqnarray}
\left\{Q^{R,+},Q^{R,-}\right\}_{\rm P}&=&-Q^{R,2}\,, \nonumber \\
\left\{Q^{R,+},Q^{R,2}\right\}_{\rm P}&=&-Q^{R,+}\cosh\left(\frac{\sqrt{C}}{2}Q^{R,-}\right)\,, \label{q-Poincare} \\
\left\{Q^{R,-},Q^{R,2}\right\}_{\rm P}&=&\frac{2}{\sqrt{C}}\sinh\left(\frac{\sqrt{C}}{2}Q^{R,-}\right)\,.  
\nonumber
\end{eqnarray}
This algebra (\ref{q-Poincare}) is not the standard $q$-deformation of $sl(2)$ \cite{Drinfeld,Jimbo}\,,  
because this is not for a deformation for the Cartan direction. 
After taking an appropriate rescaling of the charges, the algebra (\ref{q-Poincare}) is isomorphic to 
a two-dimensional $q$-deformed Poincar\'e algebra
\cite{q-Poincare,Ohn}, where the deformation parameter $q$ is defined as  
\[
q \equiv {\rm e}^{\frac{\sqrt{C}}{2}}\,. 
\]
Eventually the algebra (\ref{q-Poincare}) is also known as a non-standard $q$-deformation of $sl(2)$\,. 

\medskip 

Note that the non-local currents in (\ref{NL-right1}) are related to $j_{\mu}^{L_{+}}$ 
through a non-local map\footnote{
The non-local map has been found originally in the case of squashed sigma models \cite{KYhybrid}. }
\begin{eqnarray}
&&j^{R,2}_\mu = -2{\rm e}^{\sqrt{C}\chi}{\rm Tr}\left(T^2g^{-1}j^{L_+}_\mu g\right)\,,    
\nonumber \\ &&
j^{R,+}_\mu = -2{\rm e}^{\sqrt{C}\chi}{\rm Tr}\left(T^+g^{-1}j^{L_+}_\mu g\right)\,. 
\label{NL-map1}
\end{eqnarray}
One can directly show that the currents in (\ref{NL-map1}) are conserved. 
This implies that the left-right duality in $SL(2)$ principal chiral models is realized 
in a non-local way even after the deformation has been performed. 
The current algebra for $j^{R,a}_\mu~(a=+,-,2)$ is shown in appendix A. 

\medskip 

In fact, the current algebra (\ref{jR_algebra}) contains non-ultra local terms. 
However, note that the charges in (\ref{q-Poincare charge}) are composed of the time component 
$j^{R,a}_t(x)$ only and the associated Poisson brackets do not contain non-ultra local terms. 
Thus there is no ambiguity related to non-ultra local terms in computing the algebra (\ref{q-Poincare})\,.

\subsection{Exotic infinite-dimensional extension \label{exotic:sec}}

Our main purpose is to explore an infinite-dimensional extension of the algebra (\ref{q-Poincare}). 
That is, we would like to consider its affine extension. 

\medskip 

To find out the candidate of affine generators, let us try a non-local map to $j_{\mu}^{L_{-}}$ like 
\begin{eqnarray}
&& \widetilde{j}^{R,2}_\mu = -2{\rm e}^{-\sqrt{C}\chi}{\rm Tr}\left(T^2g^{-1}j^{L_-}_\mu g\right)\,, 
\nonumber \\  
&& \widetilde{j}^{R,+}_\mu = -2{\rm e}^{-\sqrt{C}\chi}{\rm Tr}\left(T^+g^{-1}j^{L_-}_\mu g\right)\,.   
\label{NL-map2}
\end{eqnarray}
The resulting non-local currents are given by   
\begin{eqnarray}
&&\widetilde{j}^{R,2}_\mu=-{\rm e}^{-\sqrt{C}\chi}\left(J^2_\mu-\sqrt{C}\epsilon_{\mu\nu}J^{-,\nu}\right)\,, 
\nonumber \\
&&\widetilde{j}^{R,+}_\mu=-{\rm e}^{-\sqrt{C}\chi}\left(J^+_\mu-\sqrt{C}\epsilon_{\mu\nu}J^{2,\nu}+CJ^-_\mu\right)\,, 
\end{eqnarray}
and it is an easy task to show that these are conserved.  
The corresponding conserved charges are defined as 
\begin{eqnarray}
\widetilde{Q}^{R,2}=\int^\infty_{-\infty}\!\!\!dx~\widetilde{j}^{R,2}_t(x) \equiv Q_{(1)}^{R,2}\,, \qquad
\widetilde{Q}^{R,+}=\int^\infty_{-\infty}\!\!\!dx~\widetilde{j}^{R,+}_t(x) \equiv Q_{(2)}^{R,+}\,. \label{affine}
\end{eqnarray}
The meaning of the subscripts $(1)$ and $(2)$ will be clarified later as the level of the resulting algebra.  
For convenience we rename the generators of $q$-deformed Poincar\'e algebra as 
\begin{eqnarray}
 Q_{(0)}^{R,-} \equiv Q^{R,-}\,, \quad  Q_{(0)}^{R,2} \equiv Q^{R,2}\,, \quad Q_{(0)}^{R,+} \equiv Q^{R,+}\,. \label{non-affine}
\end{eqnarray}

\medskip 

The tower structure generated by (\ref{affine}) and (\ref{non-affine}) can be deduced 
by evaluating the Poisson brackets without the knowledge on the mathematical formulation 
of the affine extension of $q$-deformed Poincar\'e algebra.  
This is a great advantage of the sigma model calculation.   

\medskip 

We shall show some examples of the Poisson brackets below. Those are enough to deduce 
the tower structure of the conserved charges\footnote{
There are subtleties in computing the algebra of conserved charges because the current algebra,  
which is shown in appendix A, contains non-ultra local terms like in the case of Yangian. 
One has to follow a possible prescription in the computation. 
This point is argued in detail in appendix B. 
}.

\medskip 

\begin{figure}[htbp]
\begin{center}
\includegraphics[scale=.5]{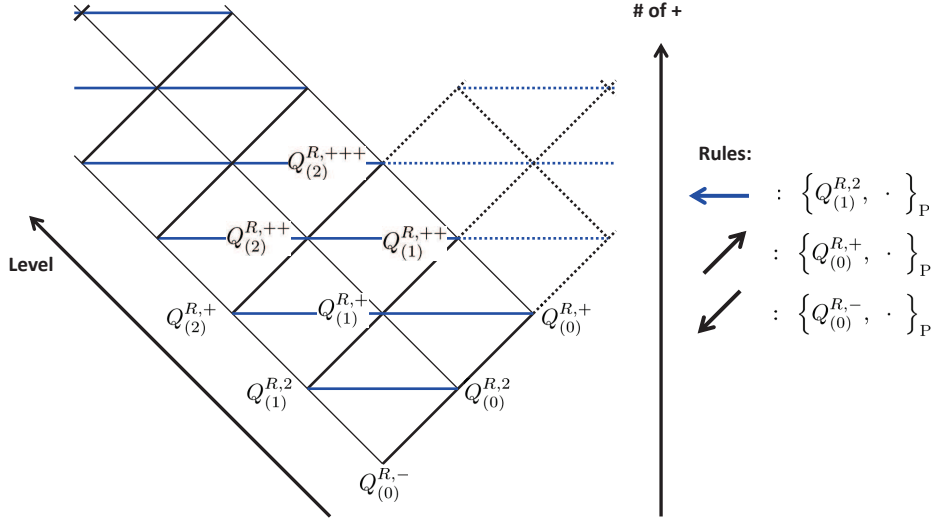} 
\end{center}
\vspace*{-3cm}
\caption{The tower structure of conserved charges. \label{tower1:fig}}  
\end{figure}

First, the Poisson bracket of $Q^{R,+}_{(0)}$ and $Q^{R,2}_{(1)}$ is evaluated as 
\begin{eqnarray}
\Bigl\{Q^{R,+}_{(0)},Q^{R,2}_{(1)}\Bigr\}_{\rm P} 
= Q_{(1)}^{R,+} + \cosh\left(\frac{\sqrt{C}}{2}Q^{R,-}_{(0)}\right)Q^{R,+}_{(0)}\,, \nonumber 
\end{eqnarray}
where a new conserved charge is defined as 
\begin{eqnarray}
Q_{(1)}^{R,+} \equiv \frac{\sqrt{C}}{2}\int^\infty_{-\infty}\!\!\!dx\int^\infty_{-\infty}\!\!\!dy~
\epsilon(x-y)\,j^{R,2}_t(x)\widetilde{j}^{R,2}_t(y)
+2\int^\infty_{-\infty}\!\!\!dx~J^+_t(x)\,.  
\end{eqnarray}
Then let us compute the following bracket, 
\begin{eqnarray}
\Bigl\{Q^{R,+}_{(0)},Q^{R,+}_{(1)}\Bigr\}_{\rm P} = Q^{R,++}_{(1)}
-\frac{\sqrt{C}}{2}\sinh\left(\frac{\sqrt{C}}{2}Q^{R,-}_{(0)}\right)Q^{R,2}_{(0)}Q^{R,+}_{(0)}\,, \label{1st}
\end{eqnarray}
where a new conserved charge $Q^{R,++}_{(1)}$ is given by 
\begin{eqnarray}
Q^{R,++}_{(1)}&\equiv &-\frac{C}{4}\int^\infty_{-\infty}\!\!\!dx\int^\infty_{-\infty}\!\!\!dy\int^\infty_{-\infty}\!\!\!dz~\epsilon(x-y)\epsilon(x-z)j^{R,2}_t(x)j^{R,2}_t(y)\widetilde{j}^{R,2}_t(z) \nonumber \\
&&+\sqrt{C}\int^\infty_{-\infty}\!\!\!dx\int^\infty_{-\infty}\!\!\!dy~\epsilon(x-y)j^{R,2}_t(x)J^+_t(y)
-2\sqrt{C}\int^\infty_{-\infty}\!\!\!dx~j^{R,+}_x(x)\,.  \label{+}
\end{eqnarray}
Note that $Q^{R,++}_{(1)}$ can also be obtained by evaluating the following bracket, 
\begin{eqnarray}
\Bigl\{Q^{R,+}_{(0)},\Bigl\{Q^{R,+}_{(0)},Q^{R,2}_{(1)}\Bigr\}_{\rm P}\Bigr\}_{\rm P} 
&=&Q^{R,++}_{(1)}
-\sqrt{C}\sinh\left(\frac{\sqrt{C}}{2}Q^{R,-}\right)Q^{R,2}Q^{R,+} \nonumber \\
&&+\frac{\sqrt{C}}{2}\sinh\left(\frac{\sqrt{C}}{2}Q^{R,-}\right)\widetilde{Q}^{R,2}Q^{R,+}
+\frac{C}{4}\left(Q^{R,2}\right)^2\widetilde{Q}^{R,2}\,. \nonumber
\end{eqnarray}
By multiplying $Q^{R,+}_{(0)}$ to $Q^{R,++}_{(1)}$\,, we obtain that 
\begin{eqnarray}
\Bigl\{Q^{R,+}_{(0)},Q^{R,++}_{(1)}\Bigr\}_{\rm P} 
&=& \frac{\sqrt{C}}{2}\sinh\left(\frac{\sqrt{C}}{2}Q^{R,-}_{(0)}\right)Q^{R,+}_{(0)}Q^{R,+}_{(1)} \nonumber \\ 
&& + \frac{C}{4}\cosh\left(\frac{\sqrt{C}}{2}Q^{R,-}_{(0)}\right)Q^{R,2}_{(0)}Q^{R,2}_{(1)}Q^{R,+}_{(0)}\,. \nonumber 
\end{eqnarray}
Hence any new conserved charges have not been generated.  
This result can be recast into the Serre relation, 
\begin{eqnarray}
&&\Bigl\{Q^{R,+}_{(0)},\Bigl\{Q^{R,+}_{(0)},\Bigl\{Q^{R,+}_{(0)},Q^{R,2}_{(1)}\Bigr\}_{\rm P}
\Bigr\}_{\rm P} \Bigr\}_{\rm P} \label{q-Serre} \\
&=&\sqrt{C}\sinh\left(\frac{\sqrt{C}}{2}Q^{R,-}_{(0)}\right)Q^{R,+}_{(0)}
\Bigl\{Q^{R,+}_{(0)},Q^{R,2}_{(1)}\Bigr\}_{\rm P}
+\frac{C}{4}\left(Q^{R,2}_{(0)}\right)^2\Bigl\{Q^{R,+}_{(0)},Q^{R,2}_{(1)}\Bigr\}_{\rm P} \nonumber \\
&&+\frac{C}{2}\cosh\left(\frac{\sqrt{C}}{2}Q^{R,-}_{(0)}\right)\left(Q^{R,2}_{(0)}\right)^2Q^{R,+}_{(0)}
-\frac{C}{2}\cosh\left(\frac{\sqrt{C}}{2}Q^{R,-}_{(0)}\right)Q^{R,2}_{(0)}Q^{R,2}_{(1)}Q^{R,+}_{(0)} \nonumber \\
&&+\frac{\sqrt{C}}{2}\sinh\left(\frac{\sqrt{C}}{2}Q^{R,-}_{(0)}\right)
\cosh\left(\frac{\sqrt{C}}{2}Q^{R,-}_{(0)}\right)\left(Q^{R,+}_{(0)}\right)^2\,. \nonumber
\end{eqnarray}
Thus evaluating possible Poisson brackets enables us to deduce the tower structure of 
the conserved charges in Figure \ref{tower1:fig}, up to lower level conserved charges. 
The level is defined as the necessary number of $\widetilde{Q}^{R,2}\,\left(=Q_{(1)}^{R,2}\right)$ 
to construct the charges at the level from the charges of $q$-deformed Poincar\'e algebra 
(This is the definition of the level 0 charges). The number of $+$ is basically measured 
by the number of charges with the $+$ group index but the $+$ index should be formally assigned to $\sqrt{C}$ 
so as to define the number of $+$ properly.  

\medskip 

Figure \ref{tower1:fig} seems a tilted Yangian algebra, but it contains a constant deformation parameter 
and the level 0 part is given by a $q$-deformed Poincar\'e algebra rather than $SL(2,\mathbb{R})_{\rm R}$\,, 
so the resulting tower structure seems to be different from the well-known Yangian (though there might be 
an isomorphism). 
In this sense this algebra should be called {\it exotic}. 

\medskip 

Note that $\widetilde{Q}^{R,+}\,(=Q_{(2)}^{R,+}) $ looks like an affine generator at a glance from the definition 
in analogy with the squashed S$^3$ case \cite{KMY-QAA}. However, 
it is not necessary to include $\widetilde{Q}^{R,+}$ into the defining relations 
so as to generate the whole tower of conserved charges.  
Eventually it is enough to take only $Q_{(1)}^{R,2}$\,. In fact, $\widetilde{Q}^{R,+}\,(=Q^{R,+}_{(2)})$ is expressed 
in terms of lower conserved charges like 
\begin{eqnarray}
\Bigl\{Q^{R,2}_{(1)},Q^{R,+}_{(1)}\Bigr\}_{\rm P}
&=&-\widetilde{Q}^{R,+} 
-\frac{\sqrt{C}}{2}\sinh\left(\frac{\sqrt{C}}{2}Q^{R,-}_{(0)}\right)\left(Q^{R,2}_{(1)}\right)^2 \\
&&
+\frac{\sqrt{C}}{4}\sinh\left(\frac{\sqrt{C}}{2}Q^{R,-}_{(0)}\right)Q^{R,2}_{(0)}Q^{R,2}_{(1)}\,. \nonumber
\end{eqnarray}
and can be regarded as one of the level 2 conserved charges. 

\medskip 

This is the reason that we have assigned the subscript $(2)$ 
as $\widetilde{Q}^{R,+} = Q_{(2)}^{R,+}$\,. 
An interpretation why the structure of this kind happens will be explained from the viewpoint 
of Riemann spheres later. 

\medskip 

\begin{figure}[htbp]
\begin{center}
\includegraphics[scale=.5]{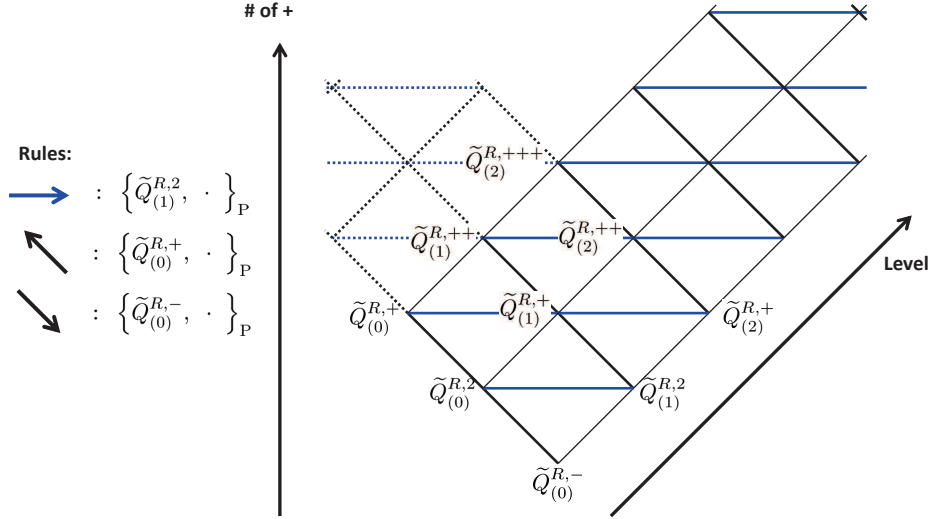} 
\end{center}
\vspace*{-3cm}
\caption{Another expression of the whole tower of conserved charges. \label{tower2:fig}}  
\end{figure}

There is another way to represent the tower of conserved charges. It is worth noting that 
the charges $\widetilde{Q}^{R,\pm}$ and $\widetilde{Q}^{R,2}$ also generate $q$-deformed Poincar\'e algebra. Hence, 
instead of $Q^{R,\pm}$ and $Q^{R,2}$\,, one can use $\widetilde{Q}^{R,\pm}$ and $\widetilde{Q}^{R,2}$ 
as the level 0 charges. 
We refer to them as $\widetilde{Q}^{R,-}_{(0)}$\,, $\widetilde{Q}^{R,2}_{(0)}$\,, and $\widetilde{Q}^{R,+}_{(0)}$\,.  
Then $Q^{R,2}$ plays a role of an affine generator in turn and should be called $\widetilde{Q}^{R,2}_{(1)}$\,.   
The tower structure is depicted in Figure \ref{tower2:fig}\,. 

\medskip 

It would probably not be difficult to notice that there is a certain relation between the two representations.  
In fact, these are exchanged through the sign flip of $\sqrt{C}$ like 
\begin{eqnarray}
\sqrt{C} \quad \longleftrightarrow \quad -\sqrt{C}\,.  \label{flipping}
\end{eqnarray}
More concretely speaking, the flipping (\ref{flipping}) exchanges the charges as 
\[
Q_{(n)}^{R,*} \quad \longleftrightarrow \quad \widetilde{Q}_{(n)}^{R,*} \qquad (* = \pm, 2)\,. 
\]
Thus the one representation is dual to the other one through (\ref{flipping})\,. 

\medskip 

Naively looking at the tower structure, 
one might think of that the two towers should be merged into a single V shape tower. 
However, it is not the case. To understand it accurately, 
it is of great value to note on the following Poisson brackets, as an example,  
\begin{eqnarray}
&&\left\{\widetilde{Q}^{R,2}_{(1)},\widetilde{Q}^{R,+}_{(0)}\right\}_{\rm P}
=-\widetilde{Q}^{R,+}_{(1)} 
-\cosh\left(\frac{\sqrt{C}}{2}\widetilde{Q}^{R,-}_{(0)}\right)\widetilde{Q}^{R,+}_{(0)}\,, \label{1} \\
&&\left\{Q^{R,2}_{(0)},Q^{R,+}_{(2)}\right\}_{\rm P}
=-Q^{R,+}_{(1)}
-\cosh\left(\frac{\sqrt{C}}{2}Q^{R,-}_{(0)}\right)Q^{R,+}_{(2)}\,. \label{2}
\end{eqnarray}
The two equations are precisely identical but \underline{only the level assignment is different.} 
The difference of the level assignment is very important 
because the interpretation of the right-hand sides of (\ref{1}) and (\ref{2}) depends on it.  

\medskip 

The first relation (\ref{1}) implies that $\widetilde{Q}^{R,+}_{(1)}$ is obtained when $\widetilde{Q}^{R,2}_{(1)}$ acts on $\widetilde{Q}^{R,+}_{(0)}$\,. 
The right-hand side of (\ref{1}) is interpreted that acting $\widetilde{Q}^{R,2}_{(1)}$ gives rise to the movement to right by one, 
because the lower conserved charges are discarded. This is depicted in Figure \ref{tower2:fig}.  

\medskip 

On the other hand, the second relation (\ref{2}) says that acting $Q^{R,2}_{(0)}$ on $Q^{R,+}_{(2)}$ does not generate any new charge. 
At a glance, it seems that the right-hand side of (\ref{2}) contains $Q^{R,+}_{(1)}$ sitting at the right of $Q^{R,+}_{(2)}$ 
and hence this operation also generates the movement to the right. 
However, $Q^{R,+}_{(1)}$ is lower than $Q^{R,+}_{(2)}$ in the level assignment in Figure \ref{tower1:fig}.  
Thus $Q^{R,+}_{(2)}$ is picked up and after all $Q^{R,+}_{(2)}$ keeps staying under the action of $Q^{R,2}_{(0)}$\,. 
In fact, any charges remain the same position under the action of $Q^{R,2}_{(0)}$ and the tower structure in Figure \ref{tower1:fig} 
is determined. This is the trick to determine the tower structure that is semi-infinite 
in a certain direction like Yangians rather than quantum affine algebras.

\medskip 

Finally we comment on the Yangian limit. When taking the $C\to 0$ limit, the $SL(2,\mathbb{R})_{\rm R}$ 
Yangian should be reproduced from the exotic symmetry discussed here. 
This limit is not obvious in the representations given in Figures \ref{tower1:fig} and \ref{tower2:fig}. 
The level 0 charges in the $SL(2,\mathbb{R})_{\rm R}$ Yangian are obtained from the charges with $2,\pm$\,. 
Then the level 1 charges are obtained from the charges with more $+$\,. For example, the level 1 charge with $+$ is obtained 
from $Q_{(1)}^{R,++}$ in (\ref{+}) by multiplying $1/\sqrt{C}$ before taking the $C\to 0$ limit. 
In fact, there is a better basis of the charges so that the Yangian limit is manifest. 
For this basis, it will be explained in \cite{future}.  

\subsection{The right Lax pairs}

It is a turn to consider Lax pairs in the right description. We refer to the Lax pairs as the right Lax pairs. 
Indeed, there are some ways to derive the right Lax pairs. Here we shall take a scaling limit of 
space-like or time-like warped AdS$_3$ spaces. Here we follow a different derivation from the one discussed 
in \cite{KY-Sch} because the argument on the left-right duality becomes much clearer. 

\medskip 

The metrics of space-like and time-like warped AdS$_3$ are given by, respectively \cite{warped},  
\begin{eqnarray}
&&ds^2=\frac{L^2}{2}\left[{\rm Tr}\left(J^2\right)-2\widetilde{C}\left({\rm Tr}\left[T^1J\right]\right)^2\right] \qquad 
(\mbox{space-like})\,, \label{w1} \\
&&ds^2=\frac{L^2}{2}\left[{\rm Tr}\left(J^2\right)-2\widetilde{C}\left({\rm Tr}\left[T^0J\right]\right)^2\right] \qquad 
(\mbox{time-like})\,. \label{w2}
\end{eqnarray}
The constant parameter $\widetilde{C}$ is a deformation parameter. 
When $\widetilde{C}=0$, the metrics (\ref{w1}) and (\ref{w2}) degenerate to the one of AdS$_3$ with radius $L$\,. 

\medskip 

Let us start from a scaling limit of the space-like warped AdS$_3$ 
to the Schr\"odinger spacetime (The process is the same for the time-like).
First, the metric (\ref{w1}) is rewritten in terms of $T^\pm$ as 
\begin{eqnarray}
ds^2=\frac{L^2}{2}\left[{\rm Tr}\left(J^2\right)-\widetilde{C}\left({\rm Tr}\left[T^-J\right]\right)^2 
+ 2\widetilde{C}{\rm Tr}\left[T^-J\right]{\rm Tr}\left[T^+J\right] 
- \widetilde{C}\left({\rm Tr}\left[T^+J\right]\right)^2\right]\,. \nonumber
\end{eqnarray}
Then $T^\pm$ is rescaled like 
\begin{eqnarray}
T^-\to\sqrt{\frac{2C}{\widetilde{C}}}\,T^-\,,\qquad T^+\to\sqrt{\frac{\widetilde{C}}{2C}}\,T^+\,.  
\end{eqnarray}
The resulting metric is given by 
\begin{eqnarray}
ds^2=\frac{L^2}{2}\left[{\rm Tr}\left(J^2\right)-2C\left({\rm Tr}\left[T^-J\right]\right)^2+2\widetilde{C}{\rm Tr}\left[T^-J\right]{\rm Tr}\left[T^+J\right]-\frac{\widetilde{C}^2}{2C}\left({\rm Tr}\left[T^+J\right]\right)^2\right]\,. \nonumber
\end{eqnarray}
Finally, by taking the following limit 
\[
\widetilde{C} \to 0  \qquad \mbox{with} ~~C~~\mbox{fixed}\,,
\]
the metric of the Schr\"odinger spacetime (\ref{last}) is reproduced. 

\medskip 

Our strategy here is to apply the same scaling limit to the Lax pairs in warped AdS$_3$ sigma models  
so as to obtain the right Lax pairs of the Schr\"odinger sigma models.  
The classical action of  two-dimensional non-linear sigma models defined 
on space-like warped AdS$_3$ is given by 
\begin{eqnarray}
S&=&-\int^\infty_{-\infty}\!\!\!dt\int^\infty_{-\infty}\!\!\!dx\,\eta^{\mu\nu}\left[{\rm Tr}\left(J_{\mu}J_{\nu}\right)
-2\widetilde{C}\,{\rm Tr}\left(T^{1}J_{\mu}\right){\rm Tr}\left(T^{1}J_{\nu}\right)\right]\,. 
\end{eqnarray}
The right Lax pair is basically given in \cite{Cherednik,FR} and its expression is
\begin{eqnarray}
&&L^R_\pm(x;\lambda_R)=-\frac{\sinh\alpha}{\sinh\left(\alpha\pm\lambda_R\right)}\left[-T^0J^0_\pm
+T^2J^2_\pm+\frac{\cosh\left(\alpha\pm\lambda_R\right)}{\cosh\alpha}T^1J^1_\pm\right]\,, 
\label{space-lax}
\\
&& \hspace*{3cm} J_\pm=J_t\pm J_x\,,\qquad \sqrt{\widetilde{C}}=\tanh\alpha\,. \nonumber 
\end{eqnarray}
Then by using the isomorphism of the $sl(2)$ algebra \cite{KMY-monodromy}
\begin{eqnarray}
\left(T^0\pm T^2\right) \to {\rm e}^{\mp\lambda_R}\left(T^0\pm T^2\right)\,, \label{iso1}
\end{eqnarray}
the Lax pair (\ref{space-lax}) is rewritten as 
\begin{eqnarray}
L^{R_+}_\pm(x;\lambda_{R_+})&=&-\frac{\sinh\alpha}{\sinh\left(\alpha\pm\lambda_{R_+}\right)}\left[
-\frac{1}{2}{\rm e}^{+\lambda_{R_+}}\left(T^0-T^2\right)\left(J^0_\pm+J^2_\pm\right) \right. 
\nonumber \\
&& \left.-\frac{1}{2}{\rm e}^{-\lambda_{R_+}}\left(T^0+T^2\right)\left(J^0_\pm-J^2_\pm\right)
+\frac{\cosh\left(\alpha\pm\lambda_{R_+}\right)}{\cosh\alpha}T^1J^1_\pm\right]\,,  \label{reduced+}
\end{eqnarray}
where $\lambda_R$ has been renamed $\lambda_{R_+}$ because the fundamental domain of $\lambda_{R_+}$ 
is now just half of the one of $\lambda_R$\,, namely the periodicity of $\lambda_{R_+}$ is $\pi$ 
while that of $\lambda_R$ is $2\pi$\,. According to this division, the number of poles contained 
in the Lax pair is also reduced from four to two. 
Thus the isomorphism (\ref{iso1}) leads to the decomposition of the Lax pair with four poles  
into a pair of the Lax pairs with two poles \cite{KMY-monodromy}.  
The Lax pair (\ref{reduced+}) is one of the Lax pairs with two poles.

\medskip 

The other Lax pair with two poles can be obtained through the $sl(2)$ isomorphism
\begin{eqnarray}
\left(T^0\pm T^2\right) \to {\rm e}^{\pm\lambda_R}\left(T^0\pm T^2\right)\,. 
\end{eqnarray}
As a result, the Lax pair is given by 
\begin{eqnarray}
L^{R_-}_\pm(x;\lambda_{R_-})&=&-\frac{\sinh\alpha}{\sinh\left(\alpha\pm\lambda_{R_-}\right)}\left[
-\frac{1}{2}{\rm e}^{-\lambda_{R_-}}\left(T^0-T^2\right)\left(J^0_\pm+J^2_\pm\right) \right. 
\label{reduced-}\\
&&\left.-\frac{1}{2}{\rm e}^{+\lambda_{R_-}}\left(T^0+T^2\right)\left(J^0_\pm-J^2_\pm\right)
+\frac{\cosh\left(\alpha\pm\lambda_{R_-}\right)}{\cosh\alpha}T^1J^1_\pm\right]\,, \nonumber 
\end{eqnarray}
where $\lambda_R$ has been renamed $\lambda_{R_-}$ on the same score. 

\medskip 

The scaling limits for (\ref{reduced+}) and (\ref{reduced-}) give rise to the following right Lax pairs, 
\begin{eqnarray}
&&L^{R_\pm}_\mu(\lambda_{R_\pm}) = \frac{1}{1-\lambda_{R_\pm}^2} 
\Bigl[-T^+\left(-J^-_\mu+\lambda_{R_\pm}\epsilon_{\mu\nu}J^{-,\nu}\right) \Bigr. \label{right-lax} \\
&& \qquad \qquad \quad \left.+T^2\left(-J^2_\mu+\lambda_{R_\pm}\epsilon_{\mu\nu}J^{2,\nu}\pm\sqrt{C}\lambda_{R_\pm} J^-_\mu\mp\sqrt{C}\lambda_{R_\pm}^2\epsilon_{\mu\nu}J^{-,\nu}\right) \right. \nonumber \\
&& \qquad \qquad \quad
\left.-T^- \Bigl( -J^+_\mu+\lambda_{R_\pm}\epsilon_{\mu\nu}J^{+,\nu}
\pm\sqrt{C}\lambda_{R_\pm} J^2_\mu\mp\sqrt{C}\lambda_{R_\pm}^2\epsilon_{\mu\nu}J^{2,\nu} \right.  \nonumber \\ 
&& \qquad \qquad \qquad \qquad 
+ C\lambda_{R_\pm} \epsilon_{\mu\nu}J^{-,\nu}-C\lambda_{R_\pm}^2J^-_\mu\Bigr)\Bigr]\,, \nonumber
\end{eqnarray}
where we have rescaled $\lambda_{R_\pm}$ as  $\lambda_{R_\pm}\to\alpha\lambda_{R_\pm}$ 
before taking the scaling limit. 
The Lax pairs (\ref{right-lax}) are also obtained by performing the following isomorphism
\begin{eqnarray}
&&T^-\to T^-\,, \nonumber \\
&&T^2 \to T^2 \mp\sqrt{C}\lambda_{R_\pm}T^-\,, \\
&&T^+ \to T^+ \mp\sqrt{C}\lambda_{R_\pm}T^2 +\frac{C}{2}\lambda_{R_\pm}^2T^-\,, \nonumber 
\end{eqnarray}
to the right Lax pair in \cite{KY-Sch}.

\medskip 

It is easy to check that the equations of motion (\ref{eom2}) are reproduced from the commutation relations 
\begin{eqnarray}
\left[\partial_t-L^{R_\pm}_t(x;\lambda_{R_\pm}),\partial_x-L^{R_\pm}_x(\lambda_{R_\pm})\right]=0\,. 
\end{eqnarray} 
Then the monodromy matrices are given by 
\begin{eqnarray}
U^{R_\pm}(\lambda_{R_\pm})={\rm P}\exp\left[\int^\infty_{-\infty}\!\!\!dx~L^{R_\pm}_x(x;\lambda_{R_\pm})\right]\,. 
\end{eqnarray}
These are conserved quantities and by expanding them with respect to $\lambda_{R_\pm}$  
an infinite set of conserved charges are obtained. By following the procedure in \cite{Maillet}\,, 
the right $r$/$s$-matrices are given by  
\begin{eqnarray}
r^{R,_\pm}(\lambda_{R,_\pm},\mu_{R,_\pm}) &=& \left(h(\lambda_{R,_\pm})+h(\mu_{R,_\pm})\right) \nonumber \\
&&\times\Biggl[\frac{1}{2\left(\lambda_{R,_\pm}-\mu_{R,_\pm}\right)}
\left(- T^+\otimes T^- - T^-\otimes T^+ + T^2\otimes T^2 \right)  
\nonumber \\
&& 
\hspace{1cm}\pm\frac{\sqrt{C}}{4}\left(T^-\otimes T^2-T^2\otimes T^-\right) \Biggr]\,, \nonumber \\
s^{R,_\pm}(\lambda_{R,_\pm},\mu_{R,_\pm}) 
&=& \left(h(\lambda_{R,_\pm})-h(\mu_{R,_\pm})\right) \nonumber \\
&&\times \Biggl[ \frac{1}{2\left(\lambda_{R,_\pm}-\mu_{R,_\pm}\right)}
\left(- T^+\otimes T^- - T^-\otimes T^+ + T^2\otimes T^2 \right)  \nonumber \\
&& \hspace{1cm}\pm\frac{\sqrt{C}}{4}\left(T^-\otimes T^2-T^2\otimes T^-\right) \Biggr]\,, \nonumber
\end{eqnarray}
with $h(\lambda)$ defined in (\ref{h-function})\,.
These satisfy the extended classical Yang-Baxter equation. 
The $r$-matrix cannot be rewritten into the skew symmetric form with respect to the difference of 
spectral parameters again.

\medskip 

Thus there are the two Lax pairs with two poles 
in the right description. Recall that there are two Lax pairs with two poles 
also in the left description. In fact, there is a relation between the right and left descriptions 
as we will see in the next subsection.

\subsection{The gauge equivalence and Riemann sphere}

So far, we have seen that there are the two descriptions 1) the left description and 2) the right description. 
Let us show that the two descriptions are related through a gauge transformation and eventually equivalent. 

\medskip  

Before arguing the gauge transformation, it is necessary to fix the relations between the spectral parameters 
in the two descriptions. These are obtained by taking a scaling limit of the spectral parameter relations, 
for example, in the case of space-like warped AdS$_3$ sigma models, 
\begin{eqnarray}
\lambda_{L_\pm}=\frac{\tanh\alpha}{\tanh\lambda_{R_\pm}} \quad\qquad \mbox{(warped~AdS$_3$)}\,. \nonumber 
\end{eqnarray}
The resulting relations after the scaling limit are simply given by  
\begin{eqnarray}
\lambda_{L_\pm}=\frac{1}{\lambda_{R_\pm}} \qquad\quad \mbox{(Schr\"odinger)}\,. \label{sp-rel}
\end{eqnarray}
This is the same as in the case of principal chiral models, up to the subscript $\pm$\,.

\medskip 

Now it is a turn to show the gauge equivalence. 
With the relations in (\ref{sp-rel})\,, the right Lax pairs are rewritten as 
\begin{eqnarray}
L^{R_\pm}_\mu(\lambda_{R_\pm}) 
&=& -\frac{1}{1-\lambda_{L_\pm}^2} 
\Bigl[-T^+\left(-\lambda_{L_\pm}^2J^-_\mu+\lambda_{L_\pm}\epsilon_{\mu\nu}J^{-,\nu}\right) \Bigr. \nonumber \\
&& \hspace*{2cm}  \left.+T^2\left(-\lambda_{L_\pm}^2J^2_\mu+\lambda_{L_\pm}\epsilon_{\mu\nu}J^{2,\nu}
\pm\sqrt{C}\lambda_{L_\pm} J^-_\mu\mp\sqrt{C}\epsilon_{\mu\nu}J^{-,\nu}\right) \right. \nonumber \\
&&  \hspace*{2cm} \Bigl.-T^- \Bigl( -\lambda_{L_\pm}^2J^+_\mu+\lambda_{L_\pm}\epsilon_{\mu\nu}J^{+,\nu}
\pm\sqrt{C}\lambda_{L_\pm} J^2_\mu\mp\sqrt{C}\epsilon_{\mu\nu}J^{2,\nu} \nonumber \\ 
&& \hspace*{2.5cm} + C\lambda_{L_\pm} \epsilon_{\mu\nu}J^{-,\nu}-CJ^-_\mu\Bigr)\Bigr]\,. \nonumber 
\end{eqnarray}
Then let us perform the gauge transformation like 
\begin{eqnarray}
&&\Bigl[L^{R_\pm}_\mu(\lambda_{R_\pm})\Bigr]^g \equiv \partial_\mu g \cdot g^{-1}+gL^{R_\pm}_\mu(\lambda_{R_\pm})g^{-1} \nonumber \\
&=&\frac{1}{1-\lambda_{L_\pm}^2} 
g\Bigl\{-T^+\left[J^-_\mu-\lambda_{L_\pm}\epsilon_{\mu\nu}J^{-,\nu}\right] \Bigr. \nonumber \\
&& \quad \left.+T^2\left[\left(J^2_\mu\pm\sqrt{C}\epsilon_{\mu\nu}J^{-,\nu}\right)
-\lambda_{L_\pm}\epsilon_{\mu\nu}\left(J^{2,\nu}\pm\sqrt{C}\epsilon^{\nu\rho}J^-_\rho\right)\right] \right. \nonumber \\
&& \quad \Bigl.-T^- \Bigl[\left(J^+_\mu\pm\sqrt{C}\epsilon_{\mu\nu}J^{2,\nu}+CJ^-_\mu\right)
-\lambda_{L_\pm}\epsilon_{\mu\nu}\left(J^{+,\nu}\pm\sqrt{C}\epsilon^{\nu\rho}J^2_\rho+CJ^{-,\nu}\right)\Bigr]\Bigr\}g^{-1} \nonumber \\
&=& L^{L_\pm}_\mu(\lambda_{L_\pm})\,. \nonumber 
\end{eqnarray}
Thus the Lax pairs in the left and right descriptions are gauge-equivalent, 
\begin{eqnarray}
\Bigl[L^{R_\pm}_\mu(\lambda_{R_\pm})\Bigr]^g  = L^{L_\pm}_\mu(\lambda_{L_\pm})\,. 
\label{gauge-eq}
\end{eqnarray}
At the level of monodromy matrix, (\ref{gauge-eq}) is recast into
\begin{eqnarray}
U^{R_{\pm}}(\lambda_{\pm})  =  g_{\infty}^{-1}\cdot U^{L_{\pm}}(\lambda_{\pm})\cdot g_{\infty}\,. 
\end{eqnarray}
Thus the right description is gauge-equivalent globally to the left one. 

\medskip 

\begin{figure}[htbp]
\begin{center}
\includegraphics[scale=.4]{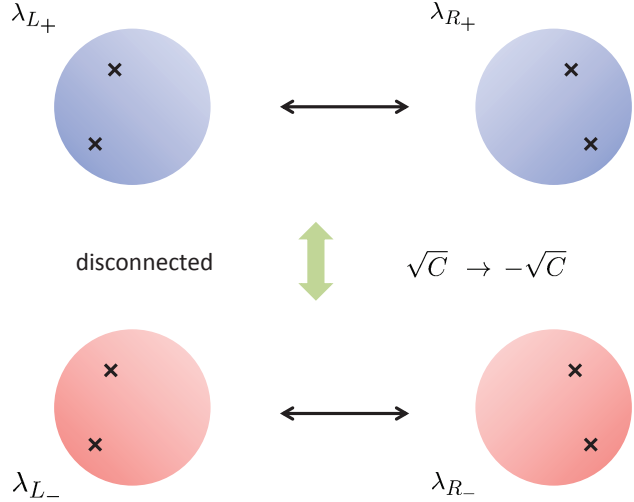}
\end{center}
\vspace*{-1cm}
\caption{The spaces of spectral parameters are described as Riemann spheres with two punctures. 
The blue spheres are gauge-equivalent to the red ones. 
\label{Riemann:fig}}
\end{figure}

We comment on the Riemann sphere picture of this equivalence as depicted in Figure \ref{Riemann:fig}.  
The pole structure of the Lax pairs is figured out from the expressions. 
For each of them, the space of spectral parameter is represented by a Riemann sphere with two punctures. 
The spectral parameter relations (\ref{sp-rel}) lead to the equivalence of the Riemann spheres of $\lambda_{R_+}$ and $\lambda_{L_+}$ 
Similar equivalence holds for $\lambda_{R_-}$ and $\lambda_{L_-}$\,. 

\medskip 

Now that we have two independent Riemann spheres with two punctures for the subscripts $+$ and $-$\,. 
Note that the two spheres are related one another through the sign flipping of $\sqrt{C}$\,.  
Then the one representation of the exotic symmetry comes from the Riemann sphere for $+$ and 
the other one comes from that for $-$\,. 

\medskip 

However, this sign flipping is equivalent to the following $sl(2)$ isomorphism and the field redefinition, 
\begin{eqnarray}
T^{\pm} ~~\to~~ - T^{\pm}\,, \qquad J^{\pm} ~~\to~~ - J^{\pm}\,. 
\end{eqnarray}
This means that the sign flipping of $\sqrt{C}$ is written as a gauge transformation.  
Thus the two Riemann spheres with $+$ and $-$ are also gauge-equivalent.  
For this reason, one of the candidate of affine generators is 
realized as one of the second level conserved charges. That is, the level 0 charges with $+$ and $2$
on the one side are realized as level 2 charges on the other side through the gauge transformation. 
Thus the two representations of the exotic symmetry are also gauge-equivalent. 
In other words, the level assignment is nothing but a gauge degree of freedom.   

\medskip 

The resulting geometry in Figure \ref{Riemann:fig} can also be interpreted 
as a scaling limit of the Riemann spheres in the case of 
warped AdS$_3$ sigma models (See \cite{KMY-monodromy}). 
Because the scaling limit contains the $\widetilde{C}\to 0$ limit, the cut shrinks to a point. 
After that, the geometry is given by Figure \ref{Riemann:fig}. 

\medskip 

Finally we shall summarize the conserved charges of infinite-dimensional symmetries and 
the expansion points of the monodromy matrices in Table \ref{list:tab}. It is easy to understand 
the expansion points for the $sl(2,{\mathbb R})_{\rm L}$ Yangian with the relations  
between the spectral parameters (\ref{sp-rel}). 
The conserved charges for the affine $q$-deformed Poincar\'e algebra can also be obtained by expanding 
the right monodromy matrices $U^{R_\pm}(\lambda_{R_\pm})$ around $\lambda_{R_{\pm}}=\infty$ \cite{future}. 
According to this, the left expansion point is automatically determined through (\ref{sp-rel})\,. 
Actually, there is a slight difference between the charges obtained in the subsection \ref{exotic:sec}
and the ones obtained from the monodromy matrices. The two sets of the charges are homomorphic 
but not always isomorphic depending on the sign of $C$\,. The details will be reported in \cite{future}.

\begin{table}[htbp]
\vspace*{0.5cm}
\begin{center}
\begin{tabular}{c||c|c}
\hline 
Charges $\setminus$ Monodromies  & $\qquad U^{L_\pm}(\lambda_{L_\pm}) \qquad$ 
& $\quad U^{R_\pm}(\lambda_{R_\pm}) \quad$ \\ 
\hline\hline 
$SL(2,{\mathbb R})_{\rm L}$ Yangian & $\infty$ & 0 \\ 
\hline
exotic symmetry
& 0 & $\infty$ \\ 
\hline\hline 
local charges & $\pm 1$ & $\pm 1$ \\ 
\hline 
\end{tabular}
\caption{The charges and the expansion points of monodromy matrices. 
\label{list:tab}}
\end{center}
\end{table}

\section{Conclusion and Discussion}

We have unveiled an {\it exotic} hidden symmetry in Schr\"odinger sigma models. 
This is an infinite-dimensional extension of $q$-deformed Poincar\'e algebra. 
We have argued the tower structure of the conserved charges by directly evaluating the Serre relations. 
The tower is generated in an {\it exotic} way and it looks like a tilted Yangian algebra. 
There are two representations of the tower structure for the exotic symmetry, 
depending on the choice of the level 0 charges, 
and the one is mapped to the other simply through the sign flipping of $\sqrt{C}$\,. 

\medskip 

There are the two descriptions to describe the classical dynamics:  
1) the left description based on $SL(2,\mathbb{R})_{\rm L}$ and 2) the right description based on $U(1)_{\rm R}$\,. 
Then the monodromy matrices in the two descriptions have been shown to be gauge-equivalent 
via the relation between the spectral parameters. Moreover, we have given 
a simple Riemann sphere interpretation of this equivalence.  
It can be understood both as the remnant of the left-right duality of 
$SL(2,{\mathbb R})$ principal chiral models and as a decoupling limit of the hybrid integrable structure 
of warped AdS$_3$ sigma models.  

\medskip 

One may suspect that there might be an isomorphism between the algebra for the exotic symmetry and 
the standard Yangian. However, at a glance, the resulting algebra looks different from the Yangian 
because it contains a $q$-deformed Poincar\'e algebra, which has a constant parameter, as the level 0 part. 
Nevertheless, there might be such an isomorphism because the gauge-equivalence of the monodromy matrices 
means that both the exotic symmetry and the Yangian are realized by expanding a single monodromy matrix. 

\medskip 

At least so far, we are not sure for the mathematical formulation of the exotic symmetry itself.  
It is important to reveal the Hopf algebraic structure of this symmetry, 
especially the coproduct. We believe that there should be a well-defined mathematical structure for it, 
not depending on the sigma model action, like in the case of non-local charges 
in $O(N)$ non-linear sigma models \cite{Luscher1,Luscher2} and Yangian \cite{Drinfeld}.

\medskip 

It is nice to look for some applications of the exotic symmetry 
in the context of warped AdS$_3$/dipole CFT$_2$ \cite{Guica,SS}. 
This duality is proposed as a toy model of the Kerr/CFT correspondence \cite{Kerr/CFT}. 
It would also be interesting to consider the quantum integrability of Schr\"odinger sigma models, 
though we have confined ourselves to the classical integrability so far. 
In this direction it is worth trying to construct the Bethe ansatz 
by following the procedure in \cite{PW,quantum1,quantum2,quantum3}. 

\medskip 

We hope that the exotic symmetry unveiled here would play a significant role 
in exploring a new aspect of integrable deformations of AdS/CFT.

\subsection*{Acknowledgments}

We would like to thank N.~Dorey, N.~MacKay, M.~Magro, V.~Regelskis and B.~Vicedo for useful discussions, 
and especially T.~Matsumoto for a collaboration at an early stage and reading the manuscript carefully. 
The work of IK was supported by the Japan Society for the Promotion of Science (JSPS). 
The work of KY was supported by the scientific grants from the Ministry of Education, Culture, Sports, Science 
and Technology (MEXT) of Japan (No.\,22740160). This work was also supported in part by the Grant-in-Aid 
for the Global COE Program ``The Next Generation of Physics, Spun 
from Universality and Emergence'' from MEXT, Japan.

\appendix

\section*{Appendix}

\section{The current algebra for $j^R_\mu$ and $\widetilde{j}^R_\mu$}

We show the Poisson brackets of the conserved currents $j^R_\mu$ and $\widetilde{j}^R_\mu$ 
that are used to compute the algebra of conserved charges in section 4. 

\medskip 

The Poisson brackets between $j^R_\mu$ are evaluated as  
\begin{eqnarray}
\Bigl\{j^{R,-}_t(x),j^{R,-}_t(y)\Bigr\}_{\rm P}&=&0\,, \nonumber \\
\Bigl\{j^{R,-}_t(x),j^{R,2}_t(y)\Bigr\}_{\rm P}&=&{\rm e}^{\sqrt{C}\chi}j^{R,-}_t(x)\delta(x-y)\,, \nonumber \\
\Bigl\{j^{R,-}_t(x),j^{R,+}_t(y)\Bigr\}_{\rm P}&=&j^{R,2}_t(x)\delta(x-y)\,, \nonumber \\
\Bigl\{j^{R,2}_t(x),j^{R,2}_t(y)\Bigr\}_{\rm P}&=&-\frac{\sqrt{C}}{2}\epsilon(x-y)\left[j^{R,2}_t(x){\rm e}^{\sqrt{C}\chi}j^{R,-}_t(y)+{\rm e}^{\sqrt{C}\chi}j^{R,-}_t(x)j^{R,2}_t(y)\right]\,, \nonumber \\
\Bigl\{j^{R,2}_t(x),j^{R,+}_t(y)\Bigr\}_{\rm P}&=&{\rm e}^{\sqrt{C}\chi}j^{R,+}_t(x)\delta(x-y) \nonumber \\
&&-\frac{\sqrt{C}}{2}\epsilon(x-y)\left[j^{R,2}_t(x)j^{R,2}_t(y)+{\rm e}^{\sqrt{C}\chi}j^{R,-}_t(x)j^{R,+}_t(y)\right]\,, \nonumber \\
\Bigl\{j^{R,+}_t(x),j^{R,+}_t(y)\Bigr\}_{\rm P}&=&-\frac{\sqrt{C}}{2}\epsilon(x-y)\left[j^{R,+}_t(x)j^{R,2}_t(y)+j^{R,2}_t(x)j^{R,+}_t(y)\right]\,, \nonumber \\
\Bigl\{j^{R,-}_t(x),j^{R,-}_x(y)\Bigr\}_{\rm P}&=&0\,, \nonumber \\
\Bigl\{j^{R,-}_t(x),j^{R,2}_x(y)\Bigr\}_{\rm P}&=&{\rm e}^{\sqrt{C}\chi}j^{R,-}_x(x)\delta(x-y)\,, \nonumber \\
\Bigl\{j^{R,-}_t(x),j^{R,+}_x(y)\Bigr\}_{\rm P}&=&j^{R,2}_x(x)\delta(x-y)+\sqrt{C}{\rm e}^{\sqrt{C}\chi}j^{R,-}_t(x)\delta(x-y) \nonumber \\ 
&& -{\rm e}^{\sqrt{C}\chi(x)}\partial_x\delta(x-y)\,, \nonumber \\
\Bigl\{j^{R,2}_t(x),j^{R,-}_x(y)\Bigr\}_{\rm P}&=&-{\rm e}^{\sqrt{C}\chi}j^{R,-}_x(x)\delta(x-y)\,, 
\label{jR_algebra}\\
\Bigl\{j^{R,2}_t(x),j^{R,2}_x(y)\Bigr\}_{\rm P}&=&-2\sqrt{C}{\rm e}^{2\sqrt{C}\chi}j^{R,-}_t(x)\delta(x-y)
+{\rm e}^{2\sqrt{C}\chi(x)}\partial_x\delta(x-y) \nonumber \\
&&-\frac{\sqrt{C}}{2}\epsilon(x-y)\left[j^{R,2}_t(x){\rm e}^{\sqrt{C}\chi}j^{R,-}_x(y)
+{\rm e}^{\sqrt{C}\chi}j^{R,-}_t(x)j^{R,2}_x(y)\right]\,, \nonumber \\
\Bigl\{j^{R,2}_t(x),j^{R,+}_x(y)\Bigr\}_{\rm P}&=&{\rm e}^{\sqrt{C}\chi}j^{R,+}_x(x)\delta(x-y) \nonumber \\
&&-\frac{\sqrt{C}}{2}\epsilon(x-y)\left[j^{R,2}_t(x)j^{R,2}_x(y)
+{\rm e}^{\sqrt{C}\chi}j^{R,-}_t(x)j^{R,+}_x(y)\right]\,, \nonumber \\
\Bigl\{j^{R,+}_t(x),j^{R,-}_x(y)\Bigr\}_{\rm P}&=&-j^{R,2}_x(x)\delta(x-y)
+\sqrt{C}{\rm e}^{\sqrt{C}\chi}j^{R,-}_t(x)\delta(x-y) \nonumber \\ 
&& -{\rm e}^{\sqrt{C}\chi(x)}\partial_x\delta(x-y)\,, \nonumber \\
\Bigl\{j^{R,+}_t(x),j^{R,2}_x(y)\Bigr\}_{\rm P}&=&-{\rm e}^{\sqrt{C}\chi}j^{R,+}_x(x)\delta(x-y) \nonumber \\
&&-\frac{\sqrt{C}}{2}\epsilon(x-y)\left[j^{R,+}_t(x){\rm e}^{\sqrt{C}\chi}
j^{R,-}_x(y)+j^{R,2}_t(x)j^{R,2}_x(y)\right]\,, \nonumber \\
\Bigl\{j^{R,+}_t(x),j^{R,+}_x(y)\Bigr\}_{\rm P}&=&-\frac{\sqrt{C}}{2}\epsilon(x-y)
\left[j^{R,+}_t(x)j^{R,2}_x(y)+j^{R,2}_t(x)j^{R,+}_x(y)\right]\,, \nonumber \\
\Bigl\{j^{R,-}_x(x),j^{R,-}_x(y)\Bigr\}_{\rm P}&=&0\,, \nonumber \\
\Bigl\{j^{R,-}_x(x),j^{R,2}_x(y)\Bigr\}_{\rm P}&=&0\,, \nonumber \\
\Bigl\{j^{R,-}_x(x),j^{R,+}_x(y)\Bigr\}_{\rm P}&=&\sqrt{C}{\rm e}^{\sqrt{C}\chi}j^{R,-}_x(x)\delta(x-y)\,, 
\nonumber \\
\Bigl\{j^{R,2}_x(x),j^{R,2}_x(y)\Bigr\}_{\rm P}&=&-\frac{\sqrt{C}}{2}\epsilon(x-y)\left[j^{R,2}_x(x){\rm e}^{\sqrt{C}\chi}j^{R,-}_x(y)+{\rm e}^{\sqrt{C}\chi}j^{R,-}_x(x)j^{R,2}_x(y)\right]\,, \nonumber \\
\Bigl\{j^{R,2}_x(x),j^{R,+}_x(y)\Bigr\}_{\rm P}&=&-\frac{\sqrt{C}}{2}\epsilon(x-y)\left[j^{R,2}_x(x)j^{R,2}_t(y) 
+{\rm e}^{\sqrt{C}\chi}j^{R,-}_x(x)j^{R,+}_t(y)\right]\,, \nonumber \\
\Bigl\{j^{R,+}_x(x),j^{R,+}_x(y)\Bigr\}_{\rm P}&=&-\frac{\sqrt{C}}{2}\epsilon(x-y)\left[j^{R,+}_x(x)j^{R,2}_x(y)+j^{R,2}_x(x)j^{R,+}_x(y)\right]\,. \nonumber
\end{eqnarray}
Note that non-ultra local terms are contained only in the Poisson brackets between $j^{R,a}_t$ and $j^{R,b}_x$\,. 
This is the same as in the case of principal chiral models. 
Thus there is no subtlety concerned with non-ultra local terms in computing (non-affine) $q$-deformed 
Poincar\'e algebra. The algebra for $\widetilde{j}^R_\mu$ can be obtained by flipping the sign of $\sqrt{C}$ 
and hence the property is the same as the above algebra.

\medskip 

Finally let us show the Poisson brackets between $j^R_\mu(x)$\,, $\widetilde{j}^R_\mu(x)$ and $J_\mu(x)$\,.
Because the complete list of the brackets is quite messy, 
we present the Poisson brackets used to discuss the exotic symmetry in section 4. 
The Poisson brackets used in this paper are the following: 
\begin{eqnarray}
\Bigl\{j^{R,2}_t(x),\widetilde{j}^{R,2}_t(y)\Bigr\}_{\rm P}&=&
-\frac{1}{2}\partial_x\left(\epsilon(x-y){\rm e}^{\sqrt{C}\chi(x)}\right)\widetilde{j}^{R,2}_t(y) \nonumber \\ 
&& -\frac{1}{2}j^{R,2}_t(x)\partial_y\left(\epsilon(x-y){\rm e}^{-\sqrt{C}\chi(y)}\right)\,, \nonumber \\
\Bigl\{j^{R,+}_t(x),\widetilde{j}^{R,2}_t(y)\Bigr\}_{\rm P}&=&\frac{\sqrt{C}}{2}\epsilon(x-y)j^{R,2}_t(x)\widetilde{j}^{R,2}_t(y)+2J^+_t(x)\delta(x-y) \\
&&-\frac{1}{2}j^{R,+}_t(x)\partial_y\left(\epsilon(x-y){\rm e}^{-\sqrt{C}\chi(y)}\right)+2\sqrt{C}\partial_x\delta(x-y)\,, \nonumber \\
\Bigl\{j^{R,+}_t(x),J^+_t(y)\Bigr\}_{\rm P}&=&-\frac{\sqrt{C}}{2}\epsilon(x-y)j^{R,+}_t(x)J^2_t(y)-\sqrt{C}j^{R,+}_x(x)\delta(x-y)\,. \nonumber \\
\Bigl\{\widetilde{j}^{R,2}_t(x),J^+_t(y)\Bigr\}_{\rm P}&=&\frac{\sqrt{C}}{2}\epsilon(x-y)\widetilde{j}^{R,2}_t(x)J^2_t(y)-\widetilde{j}^{R,+}_t(x)\delta(x-y) 
\nonumber \\ 
&& -\sqrt{C}\partial_x\left({\rm e}^{-\sqrt{C}\chi(x)}\delta(x-y)\right)\,. \nonumber
\end{eqnarray}
It is necessary for further computation to use the other brackets, as a matter of course.

\section{Prescriptions to treat non-ultra local terms}

The current algebra presented in appendix A contains non-ultra local terms. 
They lead to some ambiguities in evaluating the Poisson brackets of conserved charges. 
We here explain how to treat the ambiguities. 

\medskip

The ambiguities come from the order of limits and depend on the species of charges contained 
in the Poisson bracket. To make this point clear, let us introduce the cut-offs to the integrations  
in the level 0 charges $Q_{(0)}^R$'s and the level 1 charges $Q_{(1)}^R$'s as follows:  
\begin{eqnarray}
Q^{R,+}_{(0)}(X_1,X_2)&\equiv&\int^{X_1}_{-X_2}\!\!dx~j^{R,+}_t(x)\,, \nonumber \\
Q^{R,2}_{(1)}(X_1,X_2)&\equiv&\int^{X_1}_{-X_2}\!\!dx~\widetilde{j}^{R,2}_t(x)\,, \nonumber \\
Q_{(1)}^{R,+}(X_1,X_2)&\equiv&\frac{\sqrt{C}}{2}\int^{X_1}_{-X_2}\!\!dx\!\int^{X_1'}_{-X_2'}\!\!dx'~
\epsilon(x-x')\,j^{R,2}_t(x)\widetilde{j}^{R,2}_t(x')
+2\int^{X_1''}_{-X_2''}\!\!\!dx''~J^+_t(x'')\,, \nonumber \\
Q^{R,++}_{(1)}(X_1,X_2)&\equiv&-\frac{C}{4}\int^{X_1}_{-X_2}\!\!dx\!\int^{X_1'}_{-X_2'}\!\!dx'\!\int^{X_1''}_{-X_2''}\!\!dx''
~\epsilon(x-x')\epsilon(x-x'') j^{R,2}_t(x)j^{R,2}_t(x')\widetilde{j}^{R,2}_t(x'') \nonumber \\
&&+\sqrt{C}\int^{X_1'''}_{-X_2'''}\!\!dx'''\!\int^{X_1''''}_{-X_2''''}\!\!dx''''~\epsilon(x'''-x'''')j^{R,2}_t(x''')J^+_t(x'''') \nonumber \\
&&-2\sqrt{C}\int^{X_1'''''}_{-X_2'''''}\!\!dx'''''~j^{R,+}_x(x''''')\,. \nonumber
\end{eqnarray}
There is no ambiguity in the Poisson brackets between $Q_{(0)}^R$'s. 
Hence the first thing we consider is the bracket between $Q_{(0)}^{R}$ and $Q_{(1)}^{R}$\,. 

\medskip 

Let us, for example, see the following bracket, 
\begin{eqnarray}
&&\Bigl\{Q^{R,+}_{(0)}(X_1,X_2),Q^{R,2}_{(1)}(Y_1,Y_2)\Bigr\}_{\rm P} \nonumber \\ 
&=&  2\sqrt{C}\int^{X_1}_{-X_2}\!\!dx\!\int^{Y_1}_{-Y_2}\!\!dy~\partial_x\delta(x-y) + \mbox{(no~ambiguity)}
\nonumber \\
&=&2\sqrt{C}\left[\theta(Y_1-X_1)-\theta(Y_2-X_2)\right]  + \mbox{(no~ambiguity)}\,. \nonumber 
\end{eqnarray}
The value of  the last expression depends on the order of limits. A possible resolution of this ambiguity 
is to follow the prescription proposed by MacKay \cite{MacKay}, 
\begin{eqnarray}
X_1=X_2\equiv X\,, \qquad Y_1=Y_2\equiv Y\,. \label{prescription}
\end{eqnarray}
That is, $Q^{R,+}_{(0)}$ and $Q^{R,2}_{(1)}$ are regularized as 
\begin{eqnarray}
Q^{R,+}_{(0)}(X) \equiv \int^X_{-X}\!\!dx~j^{R,+}_t(x)\,, \qquad 
Q^{R,2}_{(1)}(Y) \equiv \int^Y_{-Y}\!\!dx~\widetilde{j}^{R,2}_t(x)\,. \nonumber
\end{eqnarray}
The prescription (\ref{prescription}) is used also in computing the Yangian algebra. 
Similarly, the other charges should be regularized following (\ref{prescription}). 
For example,  $Q^{R,+}_{(1)}$ is done as 
\begin{eqnarray}
Q_{(1)}^{R,+}(X)&\equiv&\frac{\sqrt{C}}{2}\int^X_{-X}\!\!dx\!\int^{X'}_{-X'}\!\!dx'~
\epsilon(x-x')\,j^{R,2}_t(x)\widetilde{j}^{R,2}_t(x')
+2\int^{X''}_{-X''}\!\!\!dx''~J^+_t(x'')\,. \nonumber
\end{eqnarray}

\medskip 

Next let us evaluate the Poisson bracket between $Q^{R,+}_{(0)}(X)$ and $Q^{R,+}_{(1)}(Y)$\,. 
In fact, there is another type of ambiguity. It is helpful to use the following integral formula,
\begin{eqnarray}
&&\int^b_{-b}\!\!dx~\epsilon(x-c)\left[\delta(x-a)-\delta(x+a)\right] \qquad (a>0,~b>0)\nonumber \\
&=&\int^b_c\!\!dx~\left[\delta(x-a)-\delta(x+a)\right]-\int^c_{-b}\!\!dx~\left[\delta(x-a)-\delta(x+a)\right] \nonumber \\
&=&\left[\theta(b-a)-\theta(c-a)-\theta(-c-a) 
\right]-\left[\theta(c-a)-\theta(c+a)+\theta(a-b)\right] \nonumber \\
&=&\left\{
\begin{array}{cc}
 -2\theta(c-a)+2\theta(c+a)-2\theta(a-b) & \quad (-b<c<b)  \\
 0 & {\rm otherwise}  \\
\end{array}
\right. \nonumber \\
&=&\left[\theta(c+b)-\theta(c-b)\right]\left[-2\theta(c-a)+2\theta(c+a)-2\theta(a-b)\right]\,, \label{formula}
\end{eqnarray}
where $c$ is a real constant.
With (\ref{formula}), the bracket is evaluated as follows: 
\begin{eqnarray}
&&\Bigl\{Q^{R,+}_{(0)}(X),Q^{R,+}_{(1)}(Y)\Bigr\}_{\rm P} \nonumber \\
&=& \frac{\sqrt{C}}{2}\int^X_{-X}\!\!dx\!\int^Y_{-Y}\!\!dy\!\int^{Y'}_{-Y'}\!\!dy'~\epsilon(y-y')j^{R,2}_t(y)\left(2\sqrt{C}\partial_x\delta(x-y')\right) 
+ \mbox{(no~ambiguity)}
\nonumber \\
&=& C\int^Y_{-Y}\!\!dy~j^{R,2}_t(y)\int^{Y'}_{-Y'}\!\!dy'~\epsilon(y-y')\left[\delta(y'-X)-\delta(y'+X)\right] 
+ \mbox{(no~ambiguity)}
\nonumber \\
&=&2C\int^Y_{-Y}dy~j^{R,2}_t(y)\left[\theta(y+Y')-\theta(y-Y')\right] \nonumber \\
&&\hspace{2cm}\times\left[\theta(y-X)-\theta(y+X)+\theta(X-Y')\right] + \mbox{(no~ambiguity)} \nonumber \\ 
&=&2C\int^{\min\{Y,Y'\}}_{-\min\{Y,Y'\}}dy~j^{R,2}_t(y)\left[\theta(y-X)-\theta(y+X)+\theta(X-Y')\right] \nonumber \\
&&\hspace{1cm}+ \mbox{(no~ambiguity)} \nonumber \\ 
&=&2C\theta(X-Y')\int^{\min\{Y,Y'\}}_{-\min\{Y,Y'\}}dy~j^{R,2}_t(y)-2C\int^{\min\{X,Y,Y'\}}_{-\min\{X,Y,Y'\}}dy~j^{R,2}_t(y) \nonumber \\
&&\hspace{1cm}+ \mbox{(no~ambiguity)}\,. \nonumber 
\end{eqnarray}
Thus the ambiguity of order of limits arises from the non-ultra local term. 
A possible resolution is to take the order of limits to maintain the defining relations in the mathematical formultation 
of the algebra. Indeed, the coproduct structure plays an important role in the case of Yangian \cite{MacKay}. 
However, in the present case, the mathematical formulation of the exotic symmetry has not been clarified yet, 
and the coproduct is not fixed. Hence there is no criterion so far. 

\medskip 

Here we have naively taken the $X\to \infty$ limit before the $Y'\to \infty$ limit 
where the ambiguous term vanishes, simply because the order of limits which leads to a simpler result 
works well in the case of Yangian. If the coproduct prefers the other order of limits actually, 
we have to take the additional term into account. 

\medskip 

It is a turn to evaluate the Poisson bracket between $Q^{R,+}_{(0)}$ and $Q^{R,++}_{(1)}$\,. 
Note that $Q^{R,++}_{(1)}$ is regularized as 
\begin{eqnarray}
Q^{R,++}_{(1)}(X)&\equiv&-\frac{C}{4}\int^X_{-X}\!\!dx\!\int^{X'}_{-X'}\!\!dx'\!\int^{X''}_{-X''}\!\!dx''
~\epsilon(x-x')\epsilon(x-x'')j^{R,2}_t(x)j^{R,2}_t(x')\widetilde{j}^{R,2}_t(x'') \nonumber \\
&&+\sqrt{C}\int^{X'''}_{-X'''}\!\!dx'''\!\int^{X''''}_{-X''''}\!\!dx''''~\epsilon(x'''-x'''')j^{R,2}_t(x''')J^+_t(x'''') \nonumber \\
&&-2\sqrt{C}\int^{X'''''}_{-X'''''}\!\!dx'''''~j^{R,+}_x(x''''')\,. \nonumber
\end{eqnarray}
Then, by using the formula (\ref{formula})\,, the bracket is rewritten as 
\begin{eqnarray}
&&\Bigl\{Q^{R,+}_{(0)}(X),Q^{R,++}_{(1)}(Y)\Bigr\}_{\rm P} \nonumber \\
&=&\frac{C}{4}\int^X_{-X}\!\!dx\!\int^Y_{-Y}\!\!dy\!\int^{Y'}_{-Y'}\!\!dy'\!\int^{Y''}_{-Y''}\!\!dy''
~\epsilon(y-y')\epsilon(y-y'')j^{R,2}_t(y)j^{R,2}_t(y')  \nonumber \\ 
&& \hspace*{4cm} \times \left(2\sqrt{C}\partial_x\delta(x-y'')\right) + \mbox{(no~ambiguity)}
\nonumber \\
&=&\frac{C\sqrt{C}}{2}\int^Y_{-Y}\!\!dy\!\int^{Y'}_{-Y'}\!\!dy'~\epsilon(y-y')j^{R,2}_t(y)j^{R,2}_t(y') \nonumber \\
&&\hspace{2cm}\times\int^{Y''}_{-Y''}\!\!dy''~\epsilon(y-y'')\left[\delta(y''-X)-\delta(y''+X)\right] + \mbox{(no~ambiguity)}
\nonumber \\
&=&C\sqrt{C}\int^Y_{-Y}dy\!\int^{Y'}_{-Y'}\!\!dy'~\epsilon(y-y')j^{R,2}_t(y)j^{R,2}_t(y')\left[\theta(y+Y'')-\theta(y-Y''\right]) \nonumber \\
&&\hspace{2cm}\times\left[\theta(y-X)-\theta(y+X)+\theta(X-Y'')\right]
+ \mbox{(no~ambiguity)} \nonumber \\
&=&C\sqrt{C}\int^{\min\{Y,Y''\}}_{-\min\{Y,Y''\}}dy\!\int^{Y'}_{-Y'}\!\!dy'~\epsilon(y-y')j^{R,2}_t(y)j^{R,2}_t(y') \nonumber \\
&&\hspace{2cm}\times\left[\theta(y-X)-\theta(y+X)+\theta(X-Y'')\right]
+ \mbox{(no~ambiguity)} \nonumber \\
&=& C\sqrt{C}\theta(X-Y'')\int^{\min\{Y,Y''\}}_{-\min\{Y,Y''\}}dy\!\int^{Y'}_{-Y'}\!\!dy'~\epsilon(y-y')j^{R,2}_t(y)j^{R,2}_t(y') \nonumber \\
&&-C\sqrt{C}\int^{\min\{X,Y,Y''\}}_{-\min\{X,Y,Y''\}}dy\!\int^{Y'}_{-Y'}\!\!dy'~\epsilon(y-y')j^{R,2}_t(y)j^{R,2}_t(y') 
+ \mbox{(no~ambiguity)}\,. \nonumber 
\end{eqnarray}
It seems that there is an ambiguity again because a step function is contained in the last line. 
However, the present case is a bit special.  After taking the limits $X,Y,Y',Y''\to\infty$\,,  
both terms in the last line vanish due to the symmetry, independent of the order of limits.  
Thus there is no ambiguity in this case.

\medskip 

The next task is to consider the contribution of non-ultra local terms in the Poisson brackets 
between $Q_{(1)}^R$'s. 
Let us consider, for example, the following Poisson bracket, 
\begin{eqnarray}
&&\Bigl\{Q^{R,2}_{(1)}(X),Q^{R,+}_{(1)}(Y)\Bigr\}_{\rm P} \nonumber \\
&=&2\int^X_{-X}\!\!dx\!\int^Y_{-Y}\!\!dy~\left[-\sqrt{C}\partial_x\left({\rm e}^{-\sqrt{C}\chi(x)}\delta(x-y)\right)\right]+{\rm no~ambiguity} \nonumber \\
&=&-2\sqrt{C}\,\theta(Y-X)\left[{\rm e}^{-\sqrt{C}\chi(X)}-{\rm e}^{-\sqrt{C}\chi(-X)}\right]+{\rm no~ambiguity}\,. \nonumber 
\end{eqnarray}
The term in the last line depends on the order of limits. 
In fact, if we take $X\to\infty$ before $Y\to\infty$\,, then this term vanishes.  
On the other hand, if we take $Y\to\infty$ before $X\to\infty$\,, 
the following term remains, 
\begin{eqnarray}
-2\sqrt{C}\,\theta(Y-X)\left[{\rm e}^{-\sqrt{C}\chi(X)}-{\rm e}^{-\sqrt{C}\chi(-X)}\right] ~~\rightarrow ~~
-4\sqrt{C}\,\sinh\left(\frac{\sqrt{C}}{2}Q^{R,-}_{(0)}\right)\,. \label{b3}
\end{eqnarray}
Note that there is a similar ambiguity in the case of Yangian, where a possible resolution 
is to take the order of limits so as to maintain the Serre relations \cite{MacKay}. 
However, in the present case, there is no information on the definite Serre relations. 
Although we have computed a sequence of the Poisson brackets that should correspond to 
one of the Serre relations, the invariance of it under the coproduct has not been checked yet.
After the coproduct has been revealed, the Serre relations are determined. Then, by using them, 
it would be possible to fix the ambiguity (\ref{b3})\,. 
For  the same reason as the previous, we have naively taken the former order of limits 
where the ambiguous term vanishes.

\end{document}